\documentstyle[12pt]{article}
\begin{document}
\begin{center}{\Large{\bf Noninteracting Fermions in infinite dimensions}}
\end{center}

\vskip 1.5cm

\begin{center}{\it Muktish Acharyya}\\
{\it Presidency College, 86/1 College Street,}\\
{\it Calcutta-700073, India}\\
{\it E-mail:muktish.acharyya@gmail.com}\end{center}

\vskip 2.5cm

\noindent {\bf Abstract:} Usually, we study the statistical behaviours
of noninteracting Fermions in finite (mainly two and three) dimensions.
For a fixed number of fermions, the average energy per fermion is 
calculated in two and in three dimensions and it
becomes equal to 50 and 60 per cent of the fermi energy respectively. 
However, in the higher
dimensions this percentage increases as the dimensionality increases
and in infinite dimensions it becomes 
100 per cent. This is an intersting result, at least pedagogically. 
Which implies all fermions are moving with Fermi momentum.
This result is not yet discussed in standard text books of
quantum statistics. In this paper, this fact is discussed and explained.
I hope, this article will be helpful for graduate students to study 
the behaviours of free fermions in generalised dimensionality. 

\vskip 2cm


\newpage

In the subject of quantum statistical 
mechanics, the behaviours of noninteracting fermions (or fermi gas) is
studied widely. In the text books \cite{pathria,reif,huang,schroeder},
 the fermi energy, average energy 
per fermions etc. at zero temperature ($T=0$) are discussed 
in details. Here, the 
noninteracting fermionic system is considered in three dimensions to
represent the realistic situation. 
These results are generally used to explain
a few electronic properties of the solids and usually 
discussed in the university graduate curriculum.

However, for purely pedagogical interest,  we may extend our views to higher
dimensions. We may calculate the density of fermionic states in generalised
$d$-dimensions. Hence, the average energy per fermion in 
$d$-dimensions may be calculated. 
Finally, taking the limit $d \to \infty$, we may get the
results in infinite dimensions.

In this letter, we have considered a noninteracting fermionic systems in
$d$-dimensions at $T=0$. We have calculated the density of 
fermionic states and
calculated the average energy per fermion. The study is extended to
infinite dimensions. The results are justified logically.

Let us recapitulate the theoretical results, obtained 
in two and three dimensions.
In two dimensions \cite{schroeder}, the density of fermionic 
states is $g(E)$ and the
number of fermionic states between energy $E$ and $E+dE$ is
\begin{equation}
g(E)dE = 4\pi A {{m} \over {h^2}} dE,
\end{equation}

\noindent where $A$ is the area of the system, $m$ is the mass of each
fermion and $h$ is the Planck's constant.
Interestingly, it may be noted here that in two dimensions,
the fermionic density of states is uniform (independent of energy).

In three dimensions, the
number of energy states between energy $E$ and $E+dE$ is 
\begin{equation}
g(E)dE=4\pi V(2m/h^2)^{3/2}E^{1/2}dE, 
\end{equation}
\noindent where, $V$ is the volume of the system. In three dimensions,
the density of fermionic states increases as the energy increases.
At zero temperature
(i.e., $T=0$), the Fermi-Dirac distribution function reduces to the form
\begin{eqnarray}
F(E) &=& 1 {~~~~~~{\rm for~~~~~~} E \leq E_F}\nonumber\\
&=& 0 {~~~~~~{\rm for~~~~~~} E>E_F,}
\end{eqnarray}

\noindent where $E_F$ is the fermi energy, above which all states are
empty. Now, if we calculate the average energy per fermion at $T = 0$,
we have to carry out the following integration,
\begin{equation}
<E>={{1} \over {N}}{\Large \int}_0^{E_F} E g(E)F(E) dE.
\end{equation}

\noindent Where, $N (=\int g(E)F(E)dE)$ is the total number of fermions.
Using the above forms (equations (1) and (2))
of $g(E)$, this integration can be evaluated easily to get,

\begin{equation}
<E>={{1} \over {2}} E_F,
\end{equation}

\noindent in two dimensions and

\begin{equation}
<E>={{3} \over {5}} E_F,
\end{equation}

\noindent in three dimensions.
This tells that the average energy per fermion is just 50 percent
of the fermi energy in two dimensions and 60 percent in three
dimensions. 

Now let us extend these calculations in generalised $d-$dimensions. In $d$-
dimensions, the density of fermionic states has to be calculated. The
energy of a fermion can be written as \cite{pathria} (Appendix-C)

\begin{equation}
E={{1} \over {2m}}(p_1^2+p_2^2+p_3^2+.....+p_d^2).
\end{equation}

\noindent Here, the above relation represents an equation of $d$-dimensional
hypersphere (in momentum space) having radius $R=\sqrt(2mE)$. The
density of single fermionic states will be proportional
to the volume of the spherical shell 
bounded between energy $E$ and $E+dE$. This
can be calculated easily, and gives
\begin{equation}
g_d(E) dE = C(m,V) E^{{d-2} \over {2}} dE.
\end{equation}

\noindent Here, $C(m,V)$ is a mass($m$) and volume($V$) dependent constant. 
It may be noted here, that in two dimensions,
(putting $d=2$ in above equation)  $g(E)$ is independent of $E$ \cite
{schroeder}. Similarly, in three dimensions ($d=3$), one gets the well
known and widely used result i.e., $g(E) \sim E^{1/2}$.  
To calculate the average energy
per fermion, the rest of the task is very simple. It requires just to carry
out the following integrals,

\begin{equation}
<E>_d = {{{\Large {\int}}_0^{E_F} g_d(E)E F(E) dE} \over
{{\Large {\int}}_0^{E_F} g_d(E) F(E) dE}} = {{d} \over {d+2}}E_F
\end{equation}

\noindent This is most general result in d-dimensions. 
One can easily check that $<E>=E_F/2$ in two dimensions \cite{schroeder}
and $<E>=3E_F/5$ in three dimensions \cite{pathria}. 
Hence, the well known results are restored.
Now the interesting
situation occurs, if we consider the following limit 
\begin{equation}
{\rm lim}_{d \to \infty} <E>_d = E_F.
\end{equation}

\noindent This implies that in infinite dimensions, 
each fermion has its energy equal to the fermi
energy! All fermions are moving with 
fermi momentum $p_F=\sqrt(2mE_F)$! This is an interesting result and
is not generally discussed in the standard textbooks of quantum statistical
mechanics.

Now, some questions will arise in readers mind to grasp these results.
Firstly, how is it possible that all fermions are lying in fermi
level ? This may be answered in the following way: Consider a fixed
number ($N$) of fermions. The fermi level represents the boundary in momentum 
space. If we try to accomodate finite number of fermions in any 
dimensions $d$, the number of fermions lying in the boundary (or surface)
increases as we increase the dimensionality. So, for a fixed number of
fermions, all the fermions will stay on the surface of fermi hypersphere,
if the space dimensionality is infinity.
Secondly, does it violate the Pauli principle ? The answer is, {\it 'NO'}. As
the dimensionality increases the number of energy states will also
increase. So, finite number of fermions can easily be accomodated
in infinite number (in $d \to \infty$ limit) 
of states without violating Pauli principle.

The physical quantities which involve the density of states will
be the function of dimensionalities. It would be interesting to 
extend the study for other physical quantities even at $T \neq 0$.

\vskip 0.4cm

\noindent {\bf Acknowledgments:} Author would like to thank Deepak Dhar
and Narayan Bandopadhyay for useful discussions.


\begin{center}{\bf References}\end{center}

\begin{enumerate}

\bibitem{pathria} R. K. Pathria, {\it Statistical Mechanics}, 
2/e, Elsevier, (2004) pp 198.

\bibitem{reif} F. Reif, {\it Fundamentals of Statistical and 
Thermal Physics}, Mc-Graw Hill, 1985.

\bibitem{huang} K. Huang, {\it Statistical Mechanics}, Wiley Eastern, 1991.

\bibitem{schroeder} D. V. Schroeder, {\it An Introduction to 
Thermal Physics}, Addison Wesley Publishing Company, San Fransisco, 
CA (1999) pp 282.
\end{enumerate}
\end{document}